**Article type: Communications**

**Title** Emergent magnetic states and tunable exchange bias at 3*d* nitride heterointerfaces

*Qiao Jin, Qinghua Zhang, He Bai, Amanda Huon, Timothy Charlton, Shengru Chen, Shan Lin, Haitao Hong, Ting Cui, Can Wang, Haizhong Guo, Lin Gu, Tao Zhu, Michael R. Fitzsimmons, Kui-juan Jin,\* Shanmin Wang,\* and Er-Jia Guo\**

Dr. Q. Jin, Dr. Q. H. Zhang, Miss. S. R. Chen, Dr. S. Lin, Miss. H. T. Hong, Miss. T. Cui, Prof. C. Wang, Prof. T. Zhu, Prof. K. J. Jin, and Prof. E. J. Guo
Beijing National Laboratory for Condensed Matter Physics and Institute of Physics, Chinese Academy of Sciences, Beijing 100190, China
E-mail: kjjin@iphy.ac.cn and ejguo@iphy.ac.cn

Dr. Q. Jin, Miss. S. R. Chen, Dr. S. Lin, Miss. H. T. Hong, Miss. T. Cui, Prof. C. Wang, Prof. T. Zhu, Prof. K. J. Jin, and Prof. E. J. Guo
Department of Physics & Center of Materials Science and Optoelectronics Engineering, University of Chinese Academy of Sciences, Beijing 100049, China
Dr. H. Bai
Spallation Neutron Source Science Center, Dongguan 523803, China
Prof. A. Huon
Department of Physics, Saint Joseph's University, Philadelphia, Pennsylvania 19131, USA
Prof. T. Charlton, and Prof. M. R. Fitzsimmons
Neutron Scattering Division, Oak Ridge National Laboratory, Oak Ridge, TN 37831, USA
Prof. C. Wang, Prof. T. Zhu, Prof. K. J. Jin, and Prof. E. J. Guo
Songshan Lake Materials Laboratory, Dongguan, Guangdong 523808, China
Prof. H. Z. Guo
Key Laboratory of Material Physics & School of Physics and Microelectronics, Zhengzhou University, Zhengzhou 450001, China
Prof. L. Gu
National Center for Electron Microscopy in Beijing and School of Materials Science and Engineering, Tsinghua University, Beijing 100084, China
Prof. M. R. Fitzsimmons
Department of Physics and Astronomy, University of Tennessee, Knoxville, TN 37996, USA
Prof. S. M. Wang
Department of Physics, Southern University of Science and Technology, Shenzhen 518055, China
E-mail: wangsm@sustech.edu.cn

**Abstract**: Interfacial magnetism stimulates the discovery of giant magnetoresistance and spin-orbital coupling across the heterointerfaces, facilitating the intimate correlation between spin transport and complex magnetic structures. Over decades, functional heterointerfaces composed of nitrides are seldomly explored due to the difficulty in synthesizing high-quality and correct composition nitride films. Here we report the fabrication of single-crystalline ferromagnetic $Fe_3N$ thin films with precisely controlled thickness. As film thickness decreasing, the magnetization deteriorates dramatically, and electronic state transits from metallic to insulating. Strikingly, the high-temperature ferromagnetism maintains in a $Fe_3N$ layer with a thickness down to 2 u. c. (~ 8 Å). The magnetoresistance exhibits a strong in-plane anisotropy and meanwhile the anomalous Hall resistance reserves its sign when $Fe_3N$ layer thickness exceeds 5 u. c. Furthermore, we observe a sizable exchange bias at the interfaces between a ferromagnetic $Fe_3N$ and an antiferromagnetic CrN. The exchange bias field and saturation moment strongly depend on the controllable bending curvature using cylinder diameter engineering (CDE) technique, implying the tunable magnetic states under lattice deformation. This work provides a guideline for exploring functional nitride films and applying their interfacial phenomena for innovative perspectives towards the practical applications.

**Main Text**

Magnetic materials possessing ultra-large remanent magnetization and small coercive field are generally useful in the magnetic recording and energy harvest industries, such as motors, generators, actuators, etc.[1-3] Typically, Fe-Co alloys and compounds containing rare-earth elements (such as Sm, Nd, etc.) are historically choices for the beforehand applications [4,5]. However, the cost for manufacturing and maintenance of such materials are too high to develop high-efficient, compact, and reliable devices for future technologies. Transition metal nitrides (TMNs) are a class of functional materials in which the nitrogen atoms are integrated into the interstitial sites of parent metals.[6-8] Therefore, TMNs exhibit extremely rich physical properties, for instance, high dielectricity,[9,10] high thermal conductivity,[11] superconductivity,[12,13] ferroelectricity,[14-16] as well as magnetism.[17-19] Except for these remarkable physical properties, the characteristics of superior hardness, corrosion resistance, and antioxidant make TMNs stable in air and are suitable for coating layers of functional devices.[1-3] Although TMNs have many advantages, the challenges in synthesizing single crystalline TMNs thin films with high crystallinity and correct chemical composition hinder the research in their intrinsic properties and intriguing interfacial phenomena. Taking iron nitrides as example, $Fe_{16}N_2$ has been theoretically predicted as one of the most promising rare-earth-free strong magnet candidates with a giant saturation moment (~ 3.5 $\mu_B$/Fe), which is 36% larger than that of a single-crystalline iron (~ 2.2 $\mu_B$/Fe).[20,21] Increasing the nitrogen content, $Fe_3N$ exhibits ferromagnetism with slightly reduced saturation moment. It holds the potential applications in the magnetic tunneling junctions (MTJs) and spin-polarized light-emitting diodes due to its ultra large spin polarization.[22-27] However, $Fe_3N$ often appears in other iron nitride forms due to the close thermally stable energy. The magnetic and electrical properties of Fe$_3$N have been greatly influenced by the impurities, chemical disorders, and nitrogen vacancies (NVs).[28] At present, most work are mainly focus on the amorphous, polycrystalline, and nanoparticle nitrides.[22-28] The sample quality hinders the deep investigation of their intrinsic properties in its thin film form. Therefore, mastering the advanced preparation method of TMN magnetic films with accurate stoichiometric ratio is particularly important for understanding the structure of objective substances and investigating the coupling mechanism between spin and other degrees of freedom. These factors are the core constraints that restrict the practical applications of this excellent magnetic materials.

Previously, most TMNs thin films were fabricated by reactive magnetron sputtering, radio-frequency nitrogen plasma-assisted molecular beam epitaxy (MBE) or metalorganic chemical vapor deposition (MOCVD).[29-31] The crystallinity is relatively poor because of the formation of polycrystalline grains and NVs. Recently, our group reported the fabrication of high-crystalline stoichiometric CrN ultrathin films using pulsed laser deposition (PLD) technique.[32-35] The specimens were prepared by laser ablating from a high-pressure synthesized stoichiometric target and compensating the NVs using *in-situ* atomic nitrogen plasma source. The CrN thin films fabricated in this manner exhibit a paramagnetic-to-antiferromagnetic transition at a temperature close to its bulk value ~ 283 K,[36,37] indicating the correct stoichiometry of high-quality CrN films. Thus, their intrinsic physical properties depending on the thickness, orientation, stacking sequence, and emergent interfacial phenomena have been explored in a controlled manner,[32-35] similar to the extensively investigated transition metal oxides.

In this work, we report the fabrication of stoichiometric single-crystalline Fe$_3$N thin films and their integration of antiferromagnetic CrN films with atomically sharp interfaces. The thickness-dependent magnetic and electrical phase transitions were observed. The ferromagnetic ground state of a Fe$_3$N film is maintained when its thickness reduces to 2 unit cells (u. c.). We observe a significant anomaly in the out-of-plane magnetic hysteresis loops depending on the film thickness and temperature, which may be attributed to the reorientation of canted spins. We note that the magnetic properties of a freestanding Fe$_3$N/CrN bilayer are extremely sensitive (~ 2680 emu/cm$^3$ per mm$^{-1}$) to the bending stress, providing an innovative strategy for designing uniaxial-strain sensors.

The nitridation of iron leads to a variety of phases, which strongly depend on the nitrogen content. Among them, ε-phase Fe$_3$N (ε-Fe$_3$N, abbreviation Fe$_3$N thereafter) exhibits interesting magnetic states and extends over a wide range of phase diagram.[38] A typical crystallized form of Fe$_3$N is present in **Figure 1**a. A succession of Fe-N layers is arranged along the *c* axis with N layers as spacers between Fe layers, resulting in an out-of-plane lattice constant of 4.38 Å. When viewed from [001] direction, Fe$_3$N consists of alternating atomic planes of Fe$^{3+}$ and N$^{3-}$ ions with a buckled hexagonal structure (Figure 1b). The atomic arrangement of Fe$_3$N is similar to that of a (0001) surface of α-Al$_2$O$_3$ (point group *R*-3*c*) (Figure 1c). The lattice constants of Fe$_3$N and α-Al$_2$O$_3$ are very close, yielding to a moderate lattice mismatch of -1.8%. The similar crystal symmetry and small lattice mismatch allow us to fabricate ε-phase Fe$_3$N films on the α-Al$_2$O$_3$ substrates epitaxially using plasma-assisted PLD technique. The highly reactive N atoms generated by a radio-frequency plasma source *in-situ* compensate the NVs during the growth. This process guarantees the correct stoichiometry of Fe$_3$N films. Microstructures around the interfaces between Fe$_3$N and α-Al$_2$O$_3$ are shown in Figure 1d, performed by scanning transmission electron microscopy (STEM) in the high-angle annular dark field (HAADF) mode. The STEM image indicates the high crystallinity of Fe$_3$N single layers, and the sample is free of obvious defects. The HAADF-STEM image confirms the epitaxial growth of Fe$_3$N on the substrates with atomically sharp interfaces. To note, we observe an ultrathin transition layer with a thickness of ~ 1 nm (marked in yellow arrows) between Fe$_3$N and α-Al$_2$O$_3$. This fact can be attributed to the accommodation of misfit strain. The inset of Figure 1d on the upper right shows the high magnified STEM image of a region in Fe$_3$N. The bright features indicate the positions of Fe atom columns, matching well with the atomic structure of single-phase ε-Fe$_3$N.[39] The light element N cannot be observed clearly in these STEM images due to the small atomic number ($Z$).

X-ray diffraction (XRD) measurements were carried out on a Panalytical MRD diffractometer equipped with a monochromator (only $K\alpha_1$ is allowed, $\lambda$ = 1.54 Å). The film thickness is controlled by counting number of laser pulse, further confirmed by X-ray reflectivity (XRR) (Figure S1). Figure 1e shows the XRD $\theta$-$2\theta$ scans for a 30-u. c.-thick Fe$_3$N on α-Al$_2$O$_3$ substrates. Only 00$l$ reflections from both films and substrates are observed, suggesting the epitaxial growth of Fe$_3$N films. Inset of Figure 1e presents a reciprocal space map (RSM) around substrates' 006 reflection. The clear Kiessig fringes up to three orders together with a narrow rocking curve at Fe$_3$N 002 reflection with full width at half maximum (FWHM) of ~ 0.03º (not shown) indicate the high crystallinity of as-grown Fe$_3$N films. We determined that the out-of-plane lattice constant of Fe$_3$N is ~ 4.38 Å, which is close to its bulk value. The relaxation of epitaxial misfit strain is attributed to the sharp structural transition at interfaces (Figure 1d). XPS analysis of Fe$_3$N films was carried out by investigating N 1$s$ (Figure 1f) and Fe 2$p$ (Figure 1g) core level spectra. The binding energy was carefully calibrated by taking C 1$s$ peak at 284.6 eV as a reference. Only one main peak centered at 397.4 eV was observed corresponding to N 1$s$ state due to the N content in Fe$_3$N. The prominent components of Fe 2$p$ peaks at 710 and 725 eV indicate hybridization between Fe and N. We notice that there are two weak pre-peaks at 706 and 720 eV, which may associate with metallic unbonded Fe termination layers on the top surface.[40]

The magnetic and electrical transport properties of Fe$_3$N films were investigated as a function of film thickness. **Figures 2**a and 2b shows the temperature dependent magnetization ($M$) and resistivity ($\rho$) of 6- and 30-u.c.-thick Fe$_3$N films, respectively. $M$-$T$ and $\rho$-$T$ curves for Fe$_3$N single films with various film thicknesses are shown in Figure S2. Thickness dependent Curie temperature ($T_C$), saturation moment ($M_S$), and $\rho_{300K}$ are summarized in Figures 2e-2g, respectively. $T_C$ of a 30-u.c.-thick Fe$_3$N film is ~ 440 K and reduces to ~ 385 K for a 2 -u.c.-thick Fe$_3$N film. Accompany to the reduction of $T_C$, $M_S$ of Fe$_3$N films reduces dramatically on decreasing the thickness below 10 u. c. (Figure 2c). Surprisingly, we find that $M_S$ of a 2-u.c.-thick Fe$_3$N film is ~ 150 emu/cm$^3$. In fact, this saturation moment at sub-nanometer thickness is much larger than those of most correlated magnetic oxide and nitride films. For the transport behavior of Fe$_3$N films, we obverse a clear metal-to-insulator transition as decreasing film thickness. For a 30-u.c.-thick Fe$_3$N single film, a metallic phase maintains at all temperatures. Please note that $\rho_{300K}$ of a 30-u.c.-thick Fe$_3$N is only 9.5 μΩ·cm, which is comparable to those of most noble metals, for instance Pt ($\rho_{300K}$ = 10.6 μΩ·cm) and Au ($\rho_{300K}$ = 2.2 μΩ·cm).[41,42] However, the Fe$_3$N films transit into an insulator when the thickness falls below 10 u. c. The low-temperature transport property of a 6-u.c.-thick Fe$_3$N film can be described using variable-range hopping (VRH) conduction mechanism with involving electron-electron interaction.[43] This is a typical electrical phase transition associated with a thickness-reduction induced dimensional confinement. For an ultrathin Fe$_3$N film, the free electrons are strongly localized; thus, the carrier density is significantly reduced, resulting in a sharp increase of $\rho$ by four orders of magnitude at low temperatures.

The magnetic easy-axis of Fe$_3$N films aligns in the film-plane, similar to most of strongly correlated thin films. The coercive field ($H_C$) increases on the increment of film thickness (Figure 2c). Figure 2d shows the magnetic hysteresis loops of 6- and 30-u.c.-thick Fe$_3$N films when $H // c$. We observe a consistent anomaly in all *M-H* loops of Fe$_3$N films with different film thickness. As sweeping magnetic fields from positive to negative values, the moment firstly reduces linearly and then increases abnormally at ~ 0.2 T. As further reducing magnetic field, the moment decreases monotonically. This behavior has seldomly been observed in other magnetic TMN thin films. To ensure the consistency, we record this anomaly in the *M-H* loops of a 30-u.c.-thick Fe$_3$N film as a function of temperature (Figure S3). The maximum moment at ~ 0.1−0.2 T as well as the coercive field continuously decreases as increasing temperature. We performed the MOKE measurements on a 30-u.c.-thick Fe$_3$N film. Figures 2h and 2i show two typical field-dependent evolutions of magnetic domain patterns captured by MOKE when $H // ab$ and $H // c$, respectively. The white areas represent the magnetic domains, with a negative magnetization ($M < 0$) created by the application of a negative field to positively saturated films with $M > 0$ (black areas). The density of effective nucleation sites in our films are relatively low, indicating a rather high homogeneity of samples. Please note that the switching field along the in-plane direction is an order of magnitude smaller than that along the out-of-plane direction. Therefore, we hypothesize that the anomaly in *M-H* loops ($H // c$) is attributed to the reorientation of Fe spins from in-plane to out-of-plane through a metastable state, where the partial magnetic domains align in the film plane during field-switching. The magnetoresistance (MR) of 6-u.c.- and 30-u.c.-thick Fe$_3$N films were measured as a function of temperature and magnetic field (Figures S4-S6). MR of both samples exhibit the butterfly hysteresis loops at small fields which are coupled with $H_C$ as comparatively shown in Figures 2c and 2d. Comparing MR at high fields indicates that both Fe$_3$N films have negative linear responses that increase systematically in magnitude with decreasing temperature. Interestingly, we observe a strong in-plane anisotropy in MR strongly depending on the current direction when $H // c$. For a 30-u.c.-thick Fe$_3$N film, the MR(*I // a*) is opposite to MR(*I // b*). However, MR trends for both current directions are similar when the thickness of Fe$_3$N films reduces to 6 u. c. We believe that the interesting magnetotransport behaviors are caused by the anisotropic spin alignments which are confined by its lattice structure. These results are consistent with the typical results for ferromagnets that possess strong in-plane anisotropy.

Furthermore, we investigated the magnetic exchange coupling at the interfaces between ferromagnetic Fe$_3$N and antiferromagnetic CrN layers. **Figure 3**a shows a HAADF-STEM image of a Fe$_3$N/CrN bilayer. The bright layer indicates the Fe$_3$N layer, while the dark layer is the CrN layer, because the scattering intensity of HAADF-STEM image is approximately proportional to the square of the atomic number and meanwhile the stacking density is higher in Fe$_3$N layer. This result reveals that both layers are flat and continuous over long lateral distances, suggesting the formation of high-quality interfaces. High-magnification image around Fe$_3$N/CrN interface (Figure 3b) and compositional EELS maps (Figures 3c-3f) obtained from the analysis of the Fe $L_{3,2}$-, Cr $L_{3,2}$-, N *K*-, and O *K*-edges signals reveal both layers that are epitaxially grown and chemically uniformed, as well as the interfaces are atomically sharp and contains negligible chemical intermixing.

The chemical and magnetization distributions across entire sample were firstly examined using polarized neutron reflection (PNR) technique. The measurements were conducted at 10 K after field-cooling at 1 T from room temperature. The specular neutron reflectivities are plotted in Figure 3g as a function of wave factor ($q = 4\pi\sin\theta_i/\lambda$) for the spin-up ($R^+$) and spin-down ($R^-$) polarized neutrons, where $\theta_i$ is the incident angle and $\lambda$ is the wavelength of incident neutrons. The solid lines are the best fits to the experimental data (open symbols). The figure of merit (FOM) of PNR fit yields a value of 1.89. Figure 3h shows the calculated spin asymmetry (SA) and its corresponding fit, from which the chemical (Figure 3i) and magnetization (Figure 3j) depth profiles can be obtained. The atomic density of $Fe_3N$ layer is larger than that of CrN layer. We find that the CrN layer separates into two parts: an interface layer and a top layer. The density of CrN interface layer is slightly smaller than its bulk value. This fact is possibly due to the strain compensation between structural dissimilar materials. The magnetic depth profile indicates the $Fe_3N$ layer exhibits a large magnetization (~ 1000 emu/cm$^3$) and deteriorates slightly when it is close to the CrN layer. We also notice that the CrN interface layer shows a small net moment of ~ 54 emu/cm$^3$ after field cooling. The existence of non-zero moment in the CrN interface layer is robust because the PNR fits with constraint of zero or negative $M$ in CrN interface layer have a large deviation from the experiment data, yielding FOM of 4.3 and 5.6, respectively. The field-cooled PNR data suggest that the CrN interface layer is magnetized due to the spins of Cr ions are pinned to those of Fe ions during the field cooling.[44,45] The magnetization integrated over the entire depth profile obtained from PNR is in good quantitative agreement with the magnetometry data.

The unique transport properties were measured across the $Fe_3N$/CrN interfaces from $[(CrN)_n/(Fe_3N)_n]_5$ ($C_n/F_n$) superlattices, where $n$ represents the number of unit cell of $Fe_3N$ and CrN layer, and five is the stacking periodicity of superlattices. Prior to the transport measurements, we performed the room-temperature PNR measurements on two representative $C_5/F_5$ and $C_{10}/F_{10}$ superlattices (Figures S7 and S8). These results demonstrate the uniformities of chemical and magnetization distributions within Fe$_3$N layers. The averaged saturation magnetization of Fe$_3$N layers in C$_5$/F$_5$ is nearly identical to those of Fe$_3$N layers in C$_{10}$/F$_{10}$. MR of C$_n$/F$_n$ superlattices are recorded by applying currents along *a* and *b* direction at various temperatures. **Figures 4**a-4f shows MRs at 10 K when *H // c*. We notice that MR does not show significant difference as a function of applied field when *n* is below 5 u. c. However, when *n* is beyond 5 u. c., the degeneracy of MR between two current directions appears. In Figures 4g and 4h, we show the contour plots of MR when *I // a* and *I // b* as a function of layer thickness and temperature. At low temperatures, the MR transits from negative to positive values at a critical thickness (*t*) and *t* reduces as further increasing the temperature. The strong temperature dependency in MR become subtle for thick layers. Furthermore, we also measured the Hall resistance ($\rho_{xy}$) of C$_n$/F$_n$ superlattices when *H // c* in order to probe the intriguing magnetic properties across the interfaces. We subtracted the ordinary Hall resistance ($R_0H$) from $\rho_{xy}$ to separate the anomalous Hall effect (AHE) from Hall resistance. Figures 4i-4n show the field dependent ($\rho_{xy}$-$R_0H$) from C$_n$/F$_n$ superlattices at 10 K. A saturation ($\rho_{xy}$-$R_0H$) appears for each C$_n$/F$_n$ superlattice above a critical field. For *n* = 2−4, the square-like open hysteresis loops are observed, but the value of ($\rho_{xy}$-$R_0H$) is negative. When *n* is above 5, we observe a sign reversal from negative to positive in ($\rho_{xy}$-$R_0H$) at 3 T. As further increasing *n*, ($\rho_{xy}$-$R_0H$) increases approximately linearly. Since the saturation magnetizations are almost identical in C$_n$/F$_n$ (*n* > 5) superlattices, we attribute that the thickness-driven AHE signal reversal possibly originates from the band topology (e. g. the number of crossing points) and is associated with *k*-space Berry curvature.[46] However, we do not observe the typical hump-like topological Hall signals that exist in other ferromagnetic metallic systems.[47,48] Figure 4o shows the ($\rho_{xy}$-$R_0H$) reduces towards zero as gradually increasing temperature for all C$_n$/F$_n$ superlattices. Temperature-driven AHE reduction can be understood by the slight Fermi level shift with respect to the avoided band crossing points. It is worth noting that the present spatial magnetic measurements could not detect a submicrometer-scale magnetic domains (e. g. chiral magnetic bubble-like or skyrmions) in $Fe_3N$ single films or $C_n/F_n$ superlattices across the critical thickness. Further measurements to identify the nanoscale magnetic domains and possible existing room-temperature skyrmions in such systems are needed using either magnetic force microscopy (MFM) [48,49] or Lorentz transmission electron microscopy (L-TEM).[50]

Besides the intriguing thickness-sensitive magnetotransport behaviors towards the spintronic applications, we further demonstrate that the exchange coupling across the $Fe_3N$/CrN interface is highly tunable to the mechanical bending stress. We fabricated the $Fe_3N$/CrN freestanding membranes using water-soluble single-crystalline sodium chloride substrates. The $Fe_3N$/CrN membranes attach firmly to the sapphire cylinders with pre-determined curvatures ranging from 0 to 0.25 $mm^{-1}$. The variations of bending curvature led to the successful tuning of lattice deformation.[51-53] We can deliver a roughly linear relationship between the bending stress and the cylinder curvature ($R$). The membranes support on cylinders with larger $R$ will suffer a larger lattice deformation. **Figure 5a** shows *M-H* loops of an as-grown $Fe_3N$/CrN membrane with $R = 0$ at 10 K. The magnetic hysteresis exhibits open loops with its center shifted to negative (positive) fields by an exchange bias field ($H_{EB}$) of ~ 115 Oe after 3 T (−3 T) field cooling from room temperature. This behavior suggests the presence of exchange coupling in the antiferromagnetic/ferromagnetic systems. As increasing $R$ from 0 to 0.25 $mm^{-1}$, $H_{EB}$ increases from 115 to 350 Oe, whereas the magnitude of $M_S$ reduces dramatically from 950 to 280 emu/$cm^2$ (Figures 5b-5d). The ratio between the bending curvatures and $M_S$ is approximately 2680 emu/$cm^3$ per $mm^{-1}$, demonstrating a significant advantage of applying such system for detection micro-deformation. Figures 5e and 5f summarize the curvature and temperature dependent $H_{EB}$ and $M_S$, respectively. We find that $H_{EB}$ reduces with increasing temperature and reaches zero value at the blocking temperature ($T_B$) ~ 150 K (Figure S9), which is consistent with the $T_N$ of antiferromagnetic CrN layers.[32-35] The transition temperature of $H_{EB}$ shifts to high temperature as increasing $R$. This strategy provides a convenient and continuous way to tune the magnetic properties of freestanding membranes. The advantage of this cylinder diameter engineering strategy lies in the uniform uniaxial lattice deformation of quasi-2D crystals and is readily applicable to tune other intriguing physical properties.

In summary, we fabricated the high-crystalline and stoichiometric $Fe_3N$ single films and heterostructures using plasma-assisted PLD technique. The intrinsic magnetic and electronic phase transitions at a critical thickness of ~ 5 u. c. were determined upon the reduced dimensionality. The anomaly in the out-of-plane magnetic hysteresis loop is attributed to the reorientation of magnetic domains during field sweeping. Furthermore, we investigated the evolution of magnetoresistance and AHE depending on the film thickness. The degeneracy in the in-plane magnetic responses and AHE signal reversal appears as gradually increasing layer thickness. We attribute such behaviors to the band topology that is strongly associated with $k$-space Berry curvature, a typical character for ferromagnets with chiral spin texture. Finally, a sizable in-plane exchange bias field of an AFM/FM heterointerface changes sign after field cooling to below the Néel temperature of CrN. We demonstrate that the magnetic property of such functional interface is highly tunable by modifying the diameter of supporting sapphire cylinder in a large range. To this end, our work on manipulation of physical properties of high-quality nitride single films and all nitride heterostructures can be extended to various similar systems. The application of rare-earth-element-free nitrides with emergent magnetic properties will stimulate combination with other low-dimensional nanomaterials.

**Materials and Methods**

**Fabrication of high-quality samples**

The Fe$_3$N thin films were fabricated by pulsed laser deposition (PLD) assisted by radio-frequency (RF) plasma source supplying highly active nitrogen atoms. A stoichiometric ceramic Fe$_3$N target was synthesized using a high-pressure reaction route from a mixed FeCl$_3$ and NaNH$_2$ powder. The high-quality target fabrication was conducted at the High-Pressure Lab of South University of Science and Technology (SUSTech). This technique was used in our previous work and the details of fabrication process for the nitride bulk ceramics or single-crystals were described in previous works. The Fe$_3$N thin films were deposited at an optimized condition that is the substrate temperature of 400 $^{\circ}$C, laser density of ~ 1 J/cm$^2$, and base pressure of 1×10$^{-8}$ Torr. The samples were cooled down to room-temperature under the irradiation of nitrogen plasma in order to compensate the nitrogen vacancies. The CrN/Fe$_3$N bilayer and superlattices were fabricated under the same experimental conditions. The layer thickness was carefully controlled by counting the number of laser pulses and further confirmed by X-ray reflectivity measurements. The number of bilayer repeats keeps the same for all superlattices in order to maintain the identical number of interfaces.

**Manipulation of bending curvatures of Fe$_3$N/CrN freestanding membranes using cylinder diameter engineering (CDE)**

The Fe$_3$N/CrN bilayers were fabricated on the single-crystalline sodium chloride substrates under the same conditions. After the growth, the Fe$_3$N/CrN bilayers were immersed into the de-ionized water to dissolve the sodium chloride substrates. Then, the bilayers float on the water surface. We used the sapphire cylinders with different curvatures (0, 0.1, and 0.25 mm$^{-1}$) to scoop up the Fe$_3$N/CrN freestanding membranes. Thus, the membranes attach to the sticks firmly. The samples were hot dried in the oven at 100 $^{\circ}$C for 2 hours. The membranes suffer the bending stress from sapphire cylinders with different curvatures. This leads to the lattice deformation and successful tuning of magnetic properties of membranes.

**Physical properties characterizations**

The crystallinity of samples was checked by in-house high-resolution four-circle X-ray diffractometer (XRD) using a Cu K$\alpha_1$ source (Panalytical MRD X'Pert 3). The microstructures of a $Fe_3N$ single layer and a $Fe_3N/CrN$ bilayer were examined using JEM ARM 200CF electron microscopy at the Institute of Physics, Chinese Academy of Sciences. We performed elemental-specific EELS mapping at the Fe $L$-, Cr $L$-, N $K$-, and O $K$-edges from the interested regions after background subtracting. The magnetic properties of samples were characterized by a SQUID equipped with a high-temperature unit. Both in-plane and out-of-plane magnetization were obtained ranging from 10 to 500 K. The exchange bias was recorded at low temperatures after the sample was field-cooled down to 10 K under magnetic fields of ± 3 T. The transport properties were performed using standard van der Pauw geometry with wire-bonding technique. The measurements were conducted using a 9T-PPMS. The ac current was kept at a minimum requirement of 1 µA to avoid the joule heating. XPS measurements were performed at the Institute of Physics, Chinese Academy of Sciences. Spectra were collected at both Fe $2p$ and N $1s$ core-level peaks at room-temperature.

**Room-temperature Magneto-optical Kerr effect (MOKE) measurements**

The MOKE measurements were performed using a commercial MOKE microscope from Evico Magnetics. The evolution of the magnetic domain structures was imaged and recorded using a quasi-static technique. The images were taken simultaneously by sweeping magnetic fields. Both in-plane and out-of-plane fields were applied during the imaging.

**Polarized Neutron Reflectivity (PNR) measurements**

The PNR experiments on the $Fe_3N/CrN$ superlattices were performed at the BL-4A of Spallation Neutron Source, Oak Ridge National Laboratory. An in-plane magnetic field of 0.5 T was applied during the measurements. The two superlattices were measured simultaneously at room temperature to avoid any arbitrary effects. The PNR measurements on a $Fe_3N/CrN$ bilayer were conducted at the Multipurpose Reflectometer at the Chinese Spallation Neutron Source. In this experiment, an in-plane magnetic field of 1 T was applied during the sample cooling. The measurements were conducted at the same magnetic field at 10 K. In both PNR measurements, the specular reflectivities were recorded as a function of the wave factor transfer along the film surface normal. $R^+$ and $R^-$ reflect the specular reflectivities from spin-up and spin-down polarized neutrons, respectively. The nuclear and magnetic scattering length densities (SLD) were obtained simultaneously by fitting the PNR data using GenX software.

**Supporting Information**

Supporting Information is available from the Wiley Online Library or from the author.

**Acknowledgements**

This work was supported by the National Key Basic Research Program of China (Grant Nos. 2020YFA0309100 and 2019YFA0308500), the National Natural Science Foundation of China (Grant Nos. 11974390, 11721404, 11874412, and 12174437), the Beijing Nova Program of Science and Technology (Grant No. Z191100001119112), the Beijing Natural Science Foundation (Grant No. 2202060), the Guangdong-Hong Kong-Macao Joint Laboratory for Neutron Scattering Science and Technology, and the Strategic Priority Research Program (B) of the Chinese Academy of Sciences (Grant No. XDB33030200). The PNR experiments on a CrN/Fe$_3$N bilayer were conducted at Multiple purpose Reflectometry (MR) at the Chinese Spallation Neutron Source (CSNS) and the PNR experiments on the CrN/Fe$_3$N superlattices were conducted via a user proposal at Magnetism Reflectometer (BL-4A) at the Spallation Neutron Source (SNS), a DOE Office of Science User Facility operated by Oak Ridge National Laboratory (ORNL).

**Figures and figure captions**

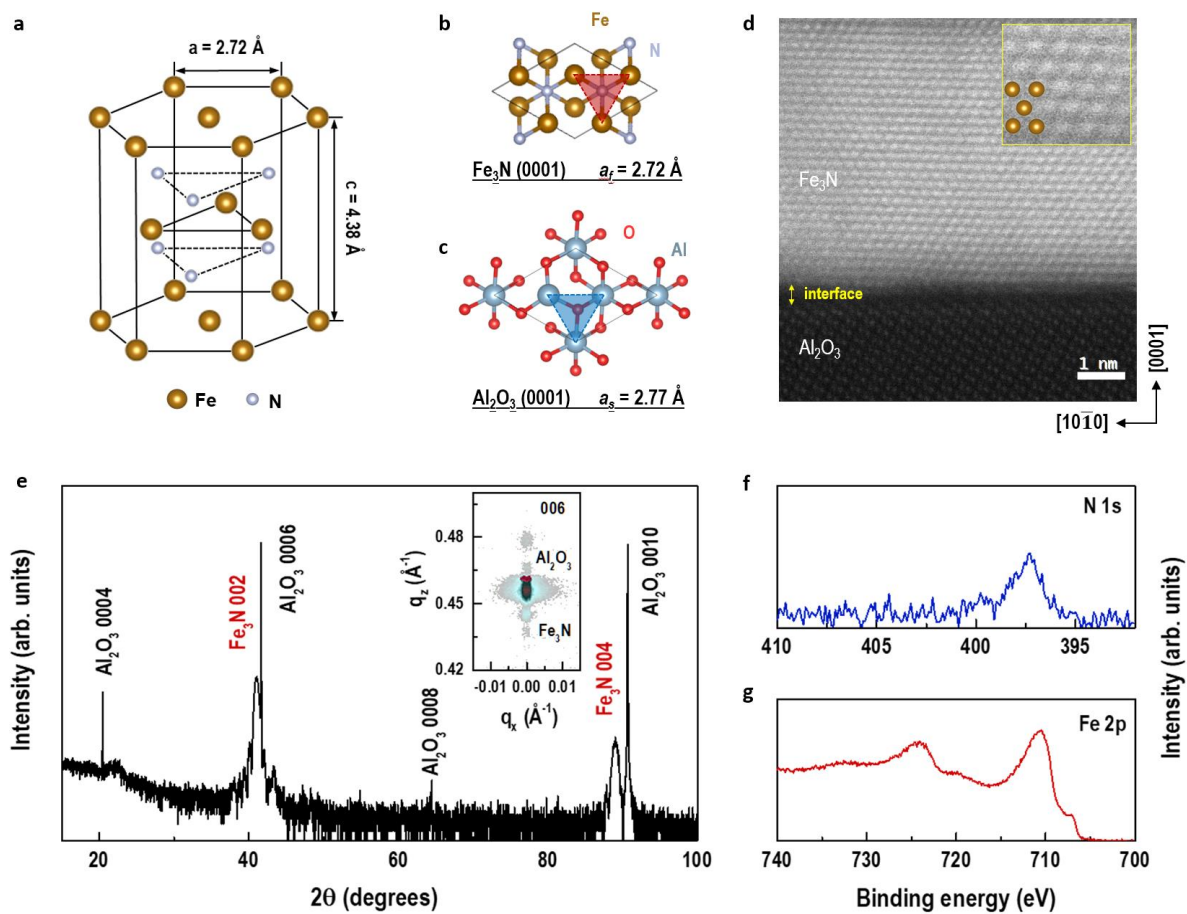

**Figure 1. Structure and electronic state characterizations of single crystalline Fe₃N films.** (a) Schematic of a Fe₃N single unit cell. (b) and (c) Top-views of Fe₃N and Al₂O₃ lattice structures, respectively, yielding to highly compatible group symmetry and relatively small misfit strain of ~ −1.8%. (d) Atomic-resolved STEM image across the Fe₃N/Al₂O₃ interface. Inset shows a representative zoom-in STEM image with atomic structures. (e) XRD $\theta$-$2\theta$ scan of a 30-u.c.-thick Fe₃N film. Inset of (e) shows the RSM around the 006 reflection of Al₂O₃ substrate. (f) and (g) N 1$s$ and Fe 2$p$ core-level XPS of a typical Fe₃N single film, confirming the presence of sufficient nitrogen content.

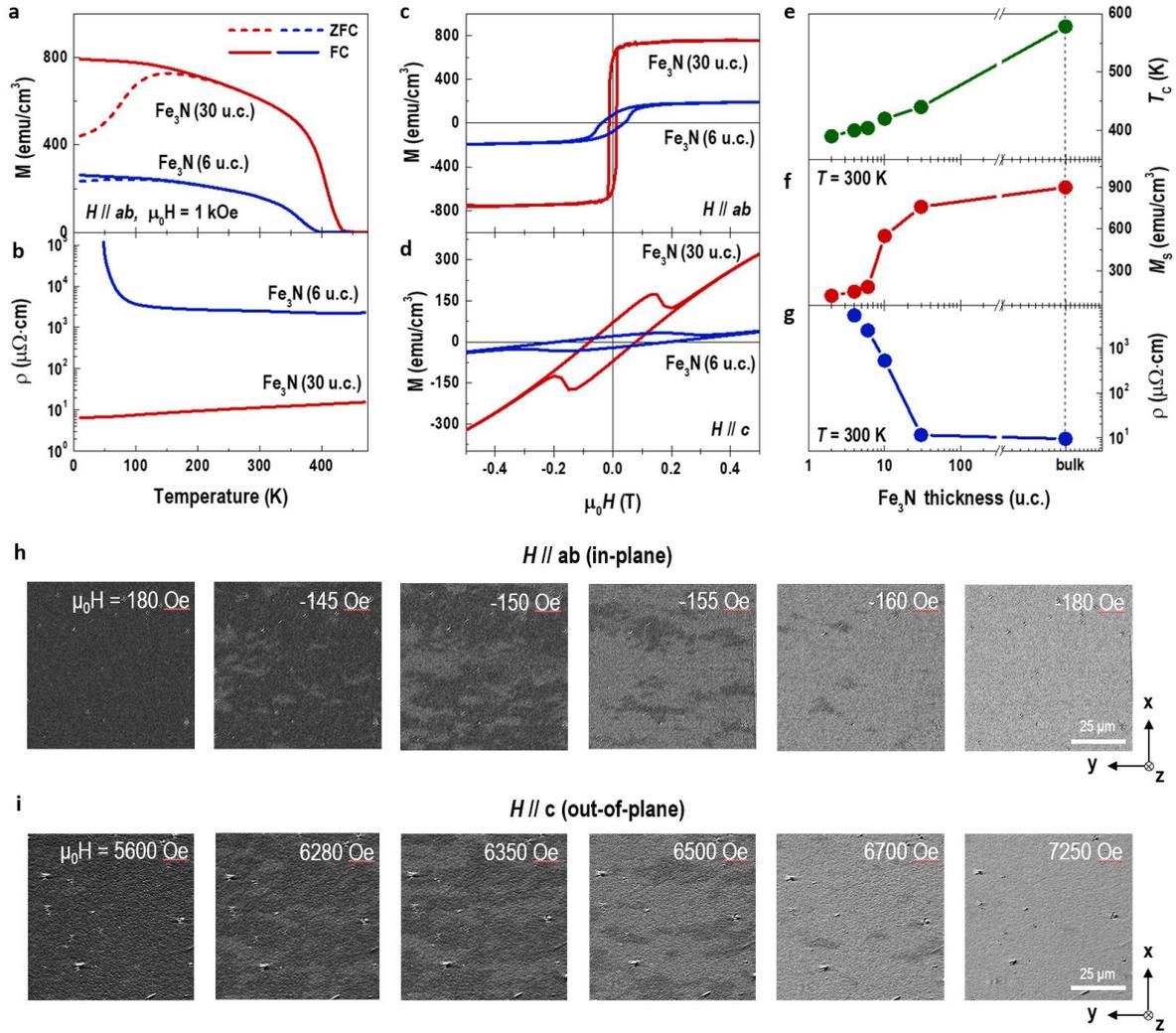

**Figure 2. Thickness dependent magnetic and transport properties of Fe₃N single films.** (a) and (b) Temperature-dependent magnetization and resistivity of 6-u.c.- and 30-u.c.-thick Fe₃N films, respectively. (c) and (d) Field-dependent magnetization of 6-u.c.- and 30-u.c.-thick Fe₃N films when magnetic field is applied along in-plane (*H*//*ab*) and out-of-plane (*H*//*c*) direction, respectively. (e) Curie temperature ($T_c$), (f) room-temperature saturation magnetization ($M_s$), and (g) resistivity ($\rho$) as a function of Fe₃N film thickness. (h) and (i) MOKE images for a 30-u.c.-thick Fe₃N film with magnetic fields applied along the in-plane and out-of-plane direction at room temperature, respectively. The dark contrast represents the magnetization along the *y* (or *z*) axis, and light contrast represents the magnetization along the -*y* (or -*z*) axis when the magnetic field is applied along the in-plane (or out-of-plane) direction.

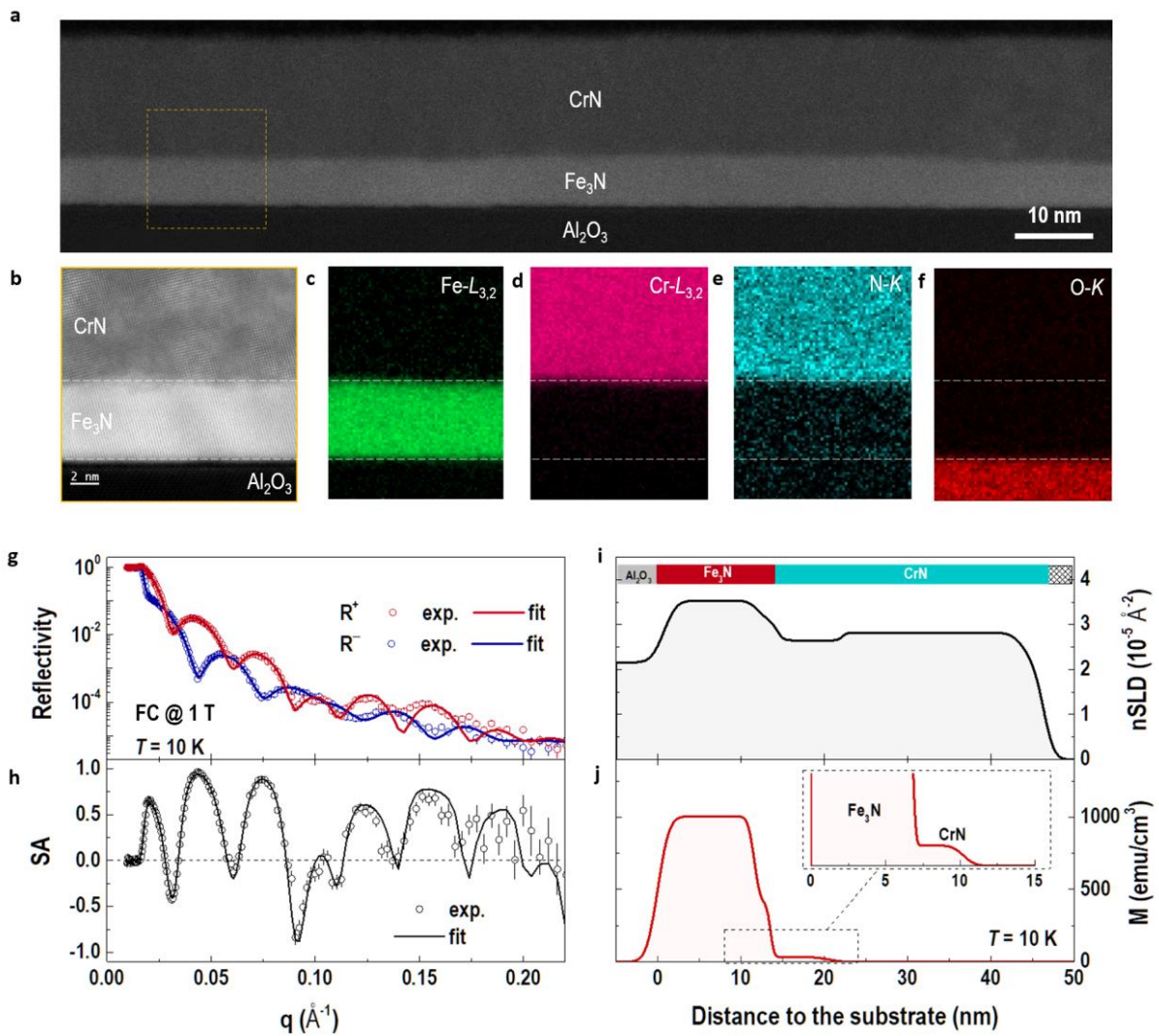

**Figure 3. Chemical and magnetization depth profiles across CrN/Fe₃N interface**. (a) Low-magnified STEM image of a CrN/Fe₃N bilayer grown on $Al_2O_3$ substrates. (b) A high-magnified STEM image across CrN/Fe₃N/$Al_2O_3$ interfaces marked with a yellow dashed square in (a). (c)-(f) Spatial resolved EELS maps were taken at the Fe $L$-, Cr $L$-, N $K$-, and O $K$-edges, respectively. The EELS results clearly indicate the uniformed chemical distribution and atomically sharp interfaces with a minimum chemical intermixing. (g) Neutron reflectivities of a CrN/Fe₃N bilayer as a function of wave factor ($q$). The open symbols (or lines) represent the reflectivities of experiment data (or best fits). (h) Calculated spin asymmetry (SA) from $(R^+ - R^-)/(R^+ + R^-)$. Open symbols and solid line are experimental data and best fit, respectively. (i) and (j) Nuclear scattering length density (nSLD) and magnetization depth profiles, respectively. Inset of (j) shows the zoom-in magnetization depth profile across the interface between CrN and Fe₃N.

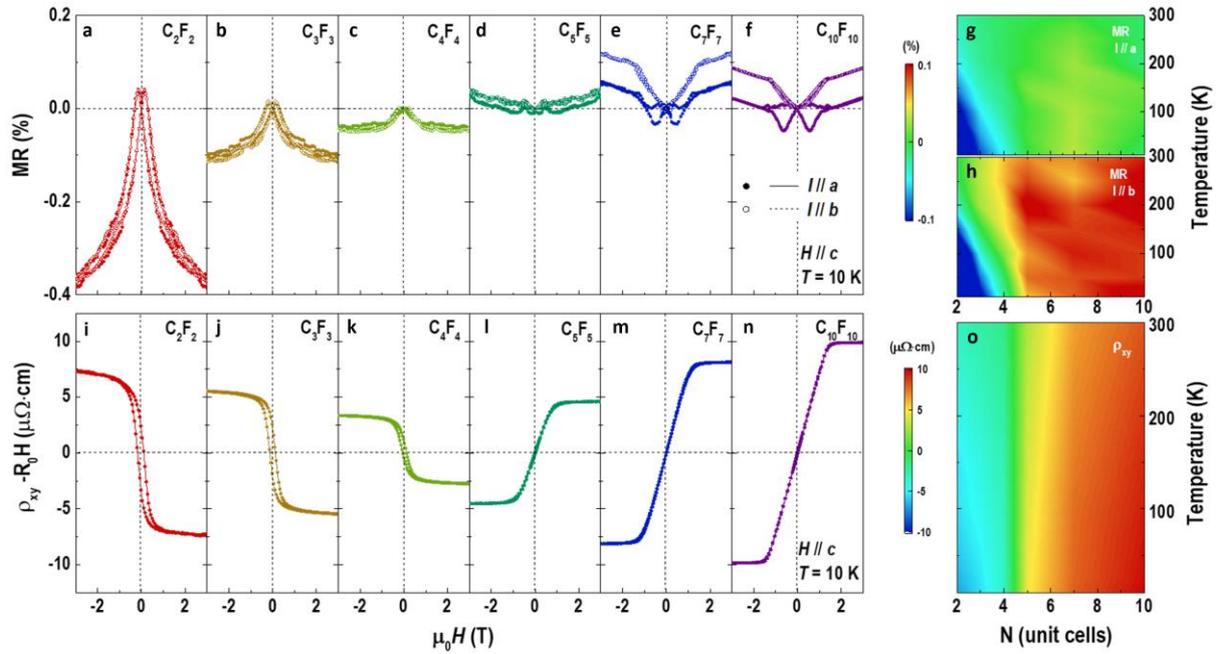

**Figure 4. Electrical transport across the [(CrN)$_n$/(Fe$_3$N)$_n$] (C$_n$F$_n$) interfaces**, where *n* represents the number of unit cells in the individual layer. (a)-(f) Magnetoresistance (MR) of C$_n$F$_n$ superlattices as a function of magnetic field at 10 K. The magnetic fields were applied perpendicular to the film plane. The currents were applied in parallel to *a* (solid symbols) and *b* (open symbols) orientations. (g) and (h) MR as a function of temperature and unit cell numbers when the current was applied in parallel to *a* and *b* orientation, respectively. (i)-(n) Anomalous Hall resistance ($\rho_{xy}$-$R_0H$) of C$_n$F$_n$ superlattices as a function of magnetic field at 10 K. (o) $\rho_{xy}$-$R_0H$ as function of temperature and unit cell numbers. For the Hall measurements, the magnetic field was applied along the surface plane.

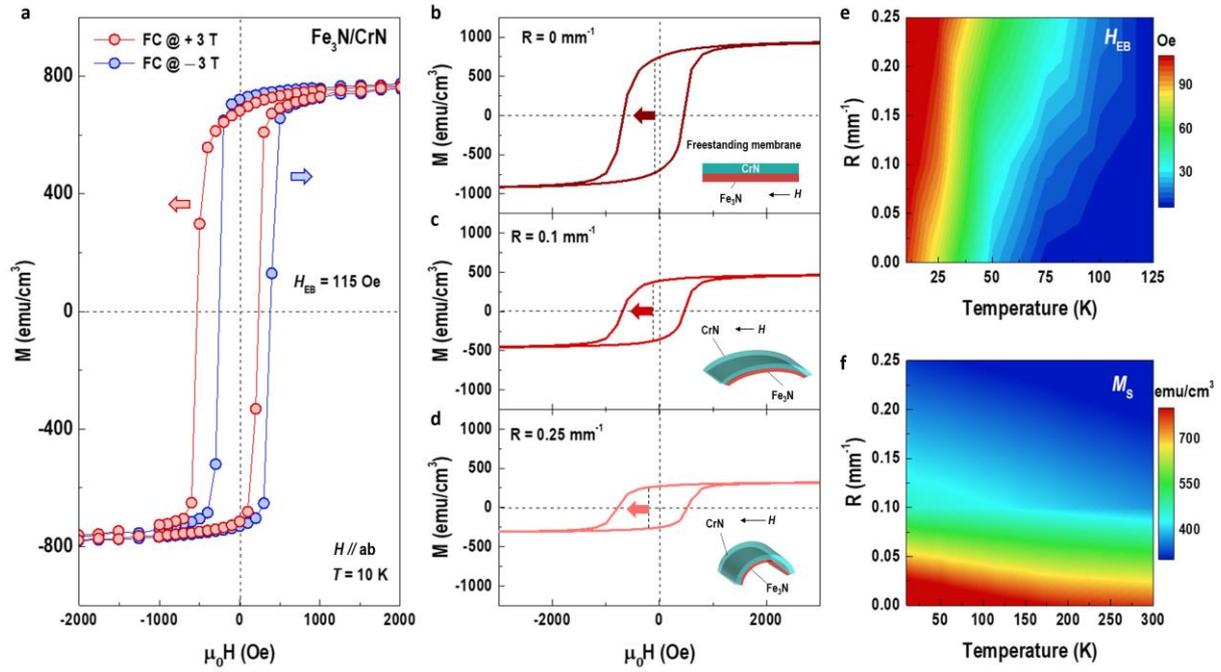

**Figure 5. Tunable magnetic states at the CrN/Fe₃N interfaces under bending stress.** (a) Exchange bias at the CrN/Fe₃N interface at 10 K. The red and blue curves are magnetic hysteresis loops after 3 and −3 T, respectively. (b)-(d) Room-temperature in-plane *M-H* loops when a freestanding CrN/Fe₃N bilayer was bended with a curvature of 0, 0.1, and 0.25 mm$^{-1}$, respectively. (e) Exchange bias field ($H_{EB}$) and (f) saturation magnetization ($M_S$) as a function of temperature and bending curvature.

25